\documentclass[
	 aps, apl
	,superscriptaddress, showpacs
	,reprint
	,amsmath,amssymb,floatfix.xcolor=table
	]{revtex4-1}

\usepackage{graphicx}
\usepackage{siunitx}
\usepackage[normalem]{ulem}
\usepackage[colorlinks=true,linkcolor=red]{hyperref}%

\usepackage{xcolor}
\newcommand\red[1]{{\color{black}#1}}

\renewcommand{\vec}[1]{\mathbf{#1}}

\newcommand{\tud}{Institut f\"{u}r Festk\"{o}per- und Materialphysik and W\"{u}rzburg-Dresden Cluster of Excellence ct.qmat, Technische Universi\"{a}t Dresden, 01062 Dresden, Germany}
\newcommand{\hzdr}{Helmholtz-Zentrum Dresden-Rossendorf e.V., Institute of Ion Beam Physics and Materials Research, 01328 Dresden, Germany}

\begin{document}

\title{Local and nonlocal spin Seebeck effect in lateral Pt-$\mathrm{Cr_2O_3}$-Pt devices at low temperatures}

\author{Prasanta Muduli}
\email{prasanta\_kumar.muduli@tu-dresden.de}
\affiliation{\tud}

\author{Richard Schlitz}
\affiliation{\tud}

\author{Tobias Kosub}
\affiliation{\hzdr}

\author{René Hübner}
\affiliation{\hzdr}

\author{Artur Erbe}
\affiliation{\hzdr}

\author{Denys Makarov}
\affiliation{\hzdr}

\author{Sebastian T. B. Goennenwein} 
\affiliation{\tud}

\red{\date{\today}}

\begin{abstract}
	 We have studied thermally driven magnon spin transport (spin Seebeck effect, SSE) in  heterostructures of antiferromagnetic $\alpha$-$\mathrm{Cr_2O_3}$ and Pt at low temperatures. Monitoring the amplitude of the local and nonlocal SSE signals as a function of temperature, we found that both decrease with increasing temperature and disappear above 100 K  and 20 K, respectively. Additionally, both SSE signals show a tendency to saturate at low temperatures. The nonlocal SSE signal decays exponentially for intermediate injector-detector separation, consistent with magnon spin current transport in the relaxation regime. 
	 We estimate the magnon relaxation length of our $\alpha$-$\mathrm{Cr_2O_3}$ films to be around 500 nm at 3 K. This short magnon relaxation length along with the strong temperature dependence of the SSE signal indicates that temperature-dependent inelastic magnon scattering processes play an important role in the intermediate range magnon transport. Our observation is relevant to low-dissipation antiferromagnetic magnon memory and logic devices involving thermal magnon generation and transport. 
\end{abstract}



\maketitle


Recently, substantial scientific effort has focused on harnessing spin currents without Joule heating for low-dissipation information processing\cite{Chumak2015}. In magnetic materials, the spin current,  (i.e., the directed propagation of spin angular momentum), can be either due to electron spin or to bosonic quasiparticle  excitations of the magnetic order parameter called magnons. The realization of magnon spin currents in electrically insulating materials with large band gap is advantageous, since they prevent  energy dissipation due to ohmic losses owing to the electronic motion. \red{Moreover, the spin propagation length of electronic spin currents is relatively short, typically ranging from nanometers in ferromagnetic metals up to micron length scales in very pure non-magnetic metals at low temperatures\cite{Bass_2007}}. Magnon spin currents, on the other hand, can travel distances up to several micrometers in magnetic insulators\cite{Kajiwara2010}. Magnons can be excited in magnetic insulators via numerous methods e.g., magnetically using microwave-frequency AC magnetic fields (coherent magnons in the GHz range), thermally via the spin Seebeck effect (incoherent magnons in the THz range)\cite{Uchida2008}, and electrically using a low-frequency AC or DC electric current in a neighbouring heavy metal with  large spin Hall angle, such as platinum\cite{Kajiwara2010}.

Although the initial study of magnon spin currents was focused on ferromagnetic insulators, recently, antiferromagnetic insulators (AFI) have moved into the focus of magnonics research due to their abundance in nature,  better scalability in nanodevices with minimal cross-talk, immunity against magnetic field perturbations, and orders of magnitude faster magnetization dynamics at the terahertz frequency range\cite{Baltz-RevModPhys.90.015005,Jungwirth2016,Jungwirth2018,Zelezny2018}.  Magnon spin current transport has been demonstrated in many antiferromagnetic insulators, such as NiO\cite{NiO-PhysRevLett.116.186601,Hou2019}, CoO\cite{Li2019-CoO}, $\mathrm{\alpha-Fe_2O_3}$\cite{Lebrun2018}, $\mathrm{MnPS_3}$\cite{MnPS3-PhysRevX.9.011026}, and $\mathrm{Cr_2O_3}$\cite{Qiu2018-Cr2O3}. Additionally, the absence of dipole-dipole interactions make axially symmetric AFI ideal materials for the realization of magnon Bose-Einstein condensates (BEC) \cite{Demokritov2006,Bozhko2016} and magnon spin superfluidity, i.e., a long-range propagating Goldstone mode arising from the spontaneous breaking of U(1) symmetry\cite{Takei-PhysRevB.90.094408,Sonin-doi:10.1080/00018731003739943,Yuaneaat1098}. All these recent studies on magnon spin transport in antiferromagnets have lead to a new frontier research field called \emph{antiferromagnetic magnonics}, as a subfield of spintronics.

One fascinating effect hinging on thermally excited magnon spin transport is the so-called spin Seebeck effect. The spin Seebeck effect refers to the generation of magnon spin currents by a temperature gradient applied across a magnetic material\cite{Uchida2008}. The SSE is manifested as an electric voltage in an adjacent heavy metal layer, in which the thermally driven spin current is converted into a charge current by the inverse spin Hall effect (ISHE)\cite{Sinova-RevModPhys.87.1213}. Originally it was assumed that AFI will not exhibit a finite SSE due to the lack of a net magnetization and the particular properties of magnon modes in AFI. More specifically, in a uniaxial AFI with two magnetic sublattices, there will be two degenerate magnon modes which produce spin current in opposite direction under a thermal gradient, such that the net spin current cancels out. However, the degeneracy of the two modes can be lifted by a magnetic field or even a spin flop transition to a ferromagnetic-like state, or by inversion symmetry breaking at an interface, resulting in finite SSE response\cite{Rezende-PhysRevB.93.014425,Bender-PhysRevLett.119.056804}. Such a magnetic field-induced SSE has been recently observed in various AFI like $\mathrm{Cr_2O_3}$\cite{Seki-PhysRevLett.115.266601,Li2020}, $\mathrm{MnF_2}$\cite{Wu-MnF2-PhysRevLett.116.097204}, and $\mathrm{FeF_2}$\cite{FeF2-PhysRevLett.122.217204}. Moreover, indication of magnon spin superfluidity has been recently reported in $\mathrm{Cr_2O_3}$\cite{Yuaneaat1098}.

In this paper, we focus on local and nonlocal SSE experiments in antiferromagnetic $\alpha$-$\mathrm{Cr_2O_3}$ thin films for small injector-detector separation. We perform magnetic-field-orientation-dependent local and nonlocal SSE measurements by rotating an external magnetic field  of constant magnitude in three orthogonal planes. By monitoring the amplitude of the local and nonlocal voltage modulation as a function of different parameters, such as the temperature  and the spatial separation between the injector-detector Pt strips, we probe the spin transport via antiferromagnetic magnons. From these data, we extract the magnon spin diffusion length ($l_m$) in the diffusive transport regime at low temperature. We show that although in the long-distance regime, superfluid spin transport might be realized, the intermediate-distance range is dominated by magnon diffusion, while the superfluid contribution is not detectable.

We use $\alpha$-$\mathrm{Cr_2O_3}$ for our spin transport experiments, since this material is one of the  best studied  AFI  with uniaxial anisotropy \cite{uniaxial-PhysRev.130.183,uniaxial2-doi:10.1080/00018735500101154}. 
The single crystal  $\alpha$-$\mathrm{Cr_2O_3}$  has hexagonal (corundum) crystal structure with $\mathrm{R\bar{3}c}$ space group. The magnetic structure is dictated by layers of $\mathrm{Cr^{3+}}$ ions (S=3/2) with an antiparallel sublattice spin alignment along the [0001] axis, such that in the (0001) plane, all $\mathrm{Cr^{3+}}$ ions belong to the same antiferromagnetic sublattice. Previous spin transport experiments have shown that the transmission of the spin current in $\mathrm{Cr_2O_3}$ depends on its Néel vector orientation\cite{Qiu2018-Cr2O3,QIN2020166362,Richard-doi:10.1063/1.5019934}. The spin transmission is completely blocked below the Néel temperature ($T_N$), if Néel vector $\vec{N}$ of $\mathrm{Cr_2O_3}$ (oriented along [0001] easy axis) is aligned perpendicular to the polarization of the injected spin current. Interestingly, $\mathrm{Cr_2O_3}$ abruptly becomes a good spin conductor above $T_N$ \cite{Qiu2018-Cr2O3}.


    The $\alpha$-$\mathrm{Cr_2O_3}$ thin films were grown by magnetron sputter deposition (base pressure: 10$^{-7}$ mbar; Ar sputter pressure: 10$^{-3}$ mbar; deposition rate: 0.04 nm/s) of a pure $\mathrm{Cr_2O_3}$ target material (robeko GmbH $\&$ Co. KG). To initiate the crystallization of the $\alpha$-$\mathrm{Cr_2O_3}$, single-crystalline Al$_2$O$_3$ (0001) substrates (Crystec GmbH) were heated to 850 $\si{\celsius}$ before the deposition. The $\alpha$-$\mathrm{Cr_2O_3}$ thin films were deposited at 700 $\si{\celsius}$. The samples were capped in-situ with a sputtered 3-nm-thick Pt layer at lower temperatures of $\approx$100 $\si{\celsius}$ (deposition rate 0.1 nm/s). \red{The films were found to be of high crystallinity and atomically smooth with single lattice steps in height at the boundary between plateaus, as discussed in more detail in ref.\cite{Patrick-Appe-doi:10.1021/acs.nanolett.8b04681}}. Structural characterization reveals (0001)-oriented growth, assuring an out-of-plane easy axis of the Néel vector for 250-nm-thick films. From cross-sectional transmission electron microscopy (TEM) (see Fig. 1(a)), the films were found to be granular with about 50-nm-sized columnar grains with very high crystallinity within each grain \red{(See Supplementary Material section S1 for further TEM characterization)} . \red{Note also from Fig. 1(a) that the Cr$_2$O$_3$/Pt interface has a certain morphological roughness, probably due to strain. However, we confirmed via energy-dispersive X-ray spectroscopy (EDXS) in scanning TEM (STEM) mode that there is no Pt diffusion into the Cr$_2$O$_3$ film (See supplementary Fig. S2). In our TEM and STEM-EDXS analyses, we also found no hints for surface oxidation or other signs for damage to the Cr$_2$O$_3$ surface due to Pt deposition.}

    Multiple lateral devices were fabricated on one and the same sample with electron beam lithography and Ar-ion milling \red{(See Supplementary Material section S2 for further fabrication details)}. Figure 1(b) shows a schematic of a Pt-$\mathrm{Cr_2O_3}$-Pt nonlocal device with two Pt strips, one acting as injector, the other as detector. The length  and width of the platinum strips for all devices studied here are approximately $L$ = 145 $\mu$m and $W$ = 500 nm, respectively. We prepared a series of nonlocal Pt-$\mathrm{Cr_2O_3}$-Pt devices with an edge-to-edge spacing $d_{nl}$ between the two  Pt strips varying from 0.4 to 4 $\mu$m. Local and nonlocal voltage  measurements were performed using a quasi-DC method applying a DC current I = 150 $\mu$A through one of the Pt strips and periodically reversing its direction. We measure both symmetric  $V^{s}_{loc(nl)}=(V_{loc(nl)}(I+)+V_{loc(nl)}(I-))/2$  and antisymmetric  $V^{as}_{loc(nl)}=(V_{loc(nl)}(I+)-V_{loc(nl)}(I-))/2$ voltages for both local (loc) and nonlocal (nl) measurement configurations. The antisymmetric voltage $V^{as}_{loc(nl)}$ and  symmetric voltage $V^{s}_{loc(nl)}$ are equivalent to the $1^{st}$ harmonic voltage ($V^{1f}$) and  second  harmonic voltage ($V^{2f}$) measured with an AC lock-in technique, respectively\cite{Goennenwein-doi:10.1063/1.4935074,Cornelissen2015,Giles-PhysRevB.92.224415,Lebrun2018,Althammer_2018,Ganzhorn-doi:10.1063/1.4986848}.

 \begin{figure}[tb]
	\centering
			\includegraphics{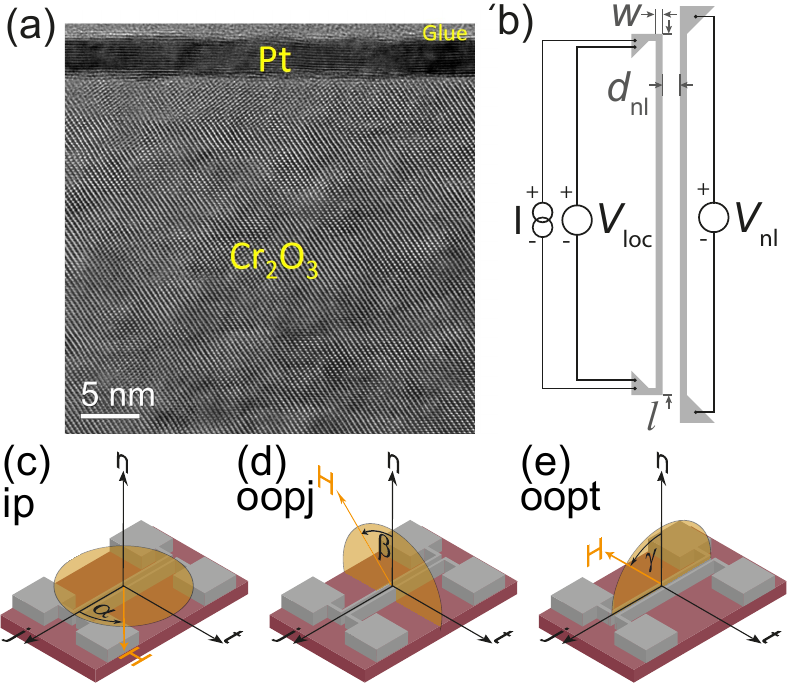} 
	\caption{%
		(a)Cross-sectional high-resolution TEM image of the $\mathrm{Cr_2O_3}$-Pt bilayer film. A high crystalline quality of the $\alpha$-$\mathrm{Cr_2O_3}$ individual grains and a sharp $\mathrm{Cr_2O_3}$-Pt interface are evident. (b) Device schematics with local and nonlocal measurement configuration. An electric current $I$ is applied at one Pt strip, and voltages can be detected at the same strip (locally) or at the other strip (nonlocally). (c,d,e) Device schematics with different external magnetic field $\vec{H}$ (yellow arrow) rotation planes:  (c) ip-rotation in the $\vec{j}-\vec{t}$ plane with angle $\alpha$ = $\angle \vec{j}\vec{H}$ ($\alpha$-rotation), (d) oopj-rotation in the $\vec{t}-\vec{n}$ plane with angle $\beta$ = $\angle \vec{n}\vec{H}$ ($\beta$-rotation), and (e) oopt-rotation in the $\vec{n}-\vec{j}$ plane with angle $\gamma$ = $\angle \vec{n}\vec{H}$ ($\gamma$-rotation).
	}
	\label{geometry}
\end{figure}

\begin{figure}[tb]
	\centering
			\includegraphics{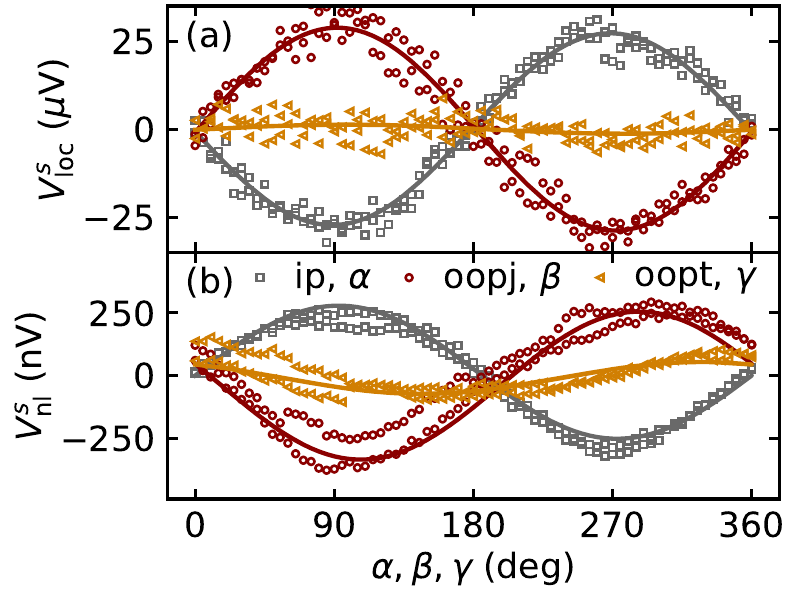} 
	\caption{%
		 Magnetic-field-orientation-dependent evolution  of the voltage symmetric in current $V^s_{loc(nl)}$ (i.e. SSE voltage) measured at 10 K \red{for the device with $d_{nl}$ = 0.4 $\mu$m}. A magnetic field of $\mu_0 H$ = 2 T was hereby used to induce a canted moment $\vec{M_{cnt}}$ in $\alpha$-$\mathrm{Cr_2O_3}$. (a) The local SSE voltage signal $V^s_{loc}$ shows a clear modulation only for $\alpha$ and $\beta$-rotations. (b) The same sin($\alpha$, $\beta$) modulation with opposite sign is observed for the nonlocal SSE voltage signal $V^s_{nl}$.  Please note, that an average offset voltage signal  was subtracted.
	}
	\label{AD-10K}
\end{figure}


 Driving a charge current through Pt  can generate magnons in the $\alpha$-$\mathrm{Cr_2O_3}$ via two different mechanisms, electrically and thermally\cite{Cornelissen2015,Shan-PhysRevB.94.174437}. Due to the spin Hall effect inside Pt,  a transverse spin current is generated orthogonal to the current direction, and a spin accumulation $\vec{\mu_s}$ with spin polarization along the \textbf{t}-axis builds up at the Pt-$\mathrm{Cr_2O_3}$ interface. When a magnetic field is applied, it can cause canting of the $\mathrm{Cr^{3+}}$ moments in the two different magnetic sublattices, producing a canted moment $\vec{M_{cnt}}$. Nonequillibrium magnon spin accumulation is created inside $\alpha$-$\mathrm{Cr_2O_3}$   via exchange interaction at the interface, if the spin accumulation direction $\vec{\mu_s}$ is not orthogonal to $\vec{M_{cnt}}$. 
Additionally, Joule heating of the injector charge current creates a temperature gradient $\nabla T$ in the $\alpha$-$\mathrm{Cr_2O_3}$ which thermally generates nonequillibrium magnons through the SSE.
The nonequillibrium magnons diffuse inside $\alpha$-$\mathrm{Cr_2O_3}$ and can be detected by the other (or the same) Pt strip. Hereby, because of the ISHE, the magnon spin current is converted into a charge current in the Pt, producing  a voltage in open-circuit condition. The electrically generated magnon voltage signal is contained in the antisymmetric part $V^{as}_{loc(nl)}$ of the measured voltage, which is linear with current\cite{Althammer_2018,Lebrun2018}. The voltage produced thermally by SSE is quadratic in current and is contained in the symmetric part $V^s_{loc(nl)}$ \red{(See Supplementary Material section S3 for further details)}. 


We measured both the antisymmetric $V^{as}_{loc(nl)}$ and the symmetric $V^{s}_{loc(nl)}$ voltages in the local and nonlocal configuration, while the external magnetic field $\vec{H}$ was rotated in three different orthogonal  planes, as depicted in Fig. 1(c,d,e). The three orthogonal rotation planes of the external magnetic field $\vec{H}$ are: (1) ip-rotation in the $\vec{j}-\vec{t}$ plane with angle $\alpha$ = $\angle \vec{j}\vec{H}$ ($\alpha$-rotation), (2) oopj-rotation in the $\vec{t}-\vec{n}$ plane with angle $\beta$ = $\angle \vec{n}\vec{H}$ ($\beta$-rotation), and (3) oopt-rotation in the $\vec{n}-\vec{j}$ plane with angle $\gamma$ = $\angle \vec{n}\vec{H}$ ($\gamma$-rotation). For the antisymmetric signal $V^{as}_{loc(nl)}$, no modulation was observed at any temperature and magnetic field \red{(See Supplementary Material section S5 for further details)}.
This can be rationalized considering that in our measurement configuration, the spin accumulation  direction $\vec{\mu_s}$ (in-plane) is perpendicular to the Néel vector $\vec{N}$ (out-of-plane). Therefore, the electrical excitation of magnons is inefficient\cite{Qiu2018-Cr2O3,QIN2020166362}. In the following, we therefore focus only on the symmetric voltage $V^s_{loc(nl)}$ which includes the thermally generated SSE signal.

Fig. 2 summarizes the angular dependence of the local and nonlocal SSE voltage for all three rotation planes at 10 K. A constant, angle-independent voltage offset was subtracted from all the data \red{(See Supplementary Material section S4 for further details)}. The angular dependence was measured applying a magnetic field $\mu_0$H  = 2 T, large enough to create a small canting of the $\mathrm{Cr^{3+}}$ moments along the field direction.  A clear modulation of the local voltage can be observed for the $\alpha$ and $\beta$-rotation planes in Fig. 2(a). The angular dependence follows a sin($\alpha$, $\beta$) dependence. This agrees with the expected behavior, for the spin Seebeck effect\cite{Uchida_2014} and thus confirms the notion that the symmetric $V^s_{loc}$ arises due to magnons excited thermally. In contrast, if the magnons were generated electrically, one would expect a sin$^2$($\alpha$, $\beta$) behavior\cite{Cornelissen2015}. 
In a simple microscopic picture, we thus assume that if the magnetic field is applied perpendicular to the [0001] easy axis (along \textbf{n}), a finite canting is induced, and consequently, a finite magnetization $\vec{M_{cnt}}$ appears in $\alpha$-$\mathrm{Cr_2O_3}$. The induced magnetization in turn gives rise to a finite spin Seebeck effect, which is maximum if \textbf{H}$\parallel$\textbf{t}, i.e.,  when the spin accumulation $\vec{\mu_s}$ direction and the induced magnetization $\vec{M_{cnt}}$ align (anti-)parallel. Consequently, the SSE vanishes, and no voltage modulation can be observed in the $\gamma$-rotation plane, where the induced magnetization $\vec{M_{cnt}}$ is always perpendicular to $\vec{\mu_s}$ (or $\vec{t}$) and thus can not generate a voltage in the given geometry (no magnons can be excited or detected).

For the nonlocal signal, a slightly different behavior is observed, as shown in Fig. 2(b). Although the signal magnitude is smaller, a clear modulation of the nonlocal SSE voltage ($V^s_{nl}$)  is observed during magnetic field rotation in the  $\alpha$ and $\beta$-plane. The signal is  similar to the local $V^s_{loc}$ and follows a sin($\alpha$, $\beta$) dependence, however, with opposite sign. \red{In thermally excited magnon transport experiments in ferrimagnetic YIG the sign of the nonlocal SSE signal reverses at a critical distance $d_{rev}$.} The critical distance $d_{rev}$ is determined by the thickness of the magnetic insulator and the interfacial spin transparency of the injector (detector) contacts\cite{Zhou-doi:10.1063/1.4976074,Shan-PhysRevB.96.184427,Shan-PhysRevB.94.174437,Ganzhorn-doi:10.1063/1.4986848}. In our case, we thus infer that the injector-detector separation $d_{nl}$ is larger than $d_{rev}$ in all the devices, since we invariably observe the same sign in $V^s_{nl}$ which is opposite to the sign of $V^s_{loc}$.


\begin{figure}[tb]
	\centering
			\includegraphics{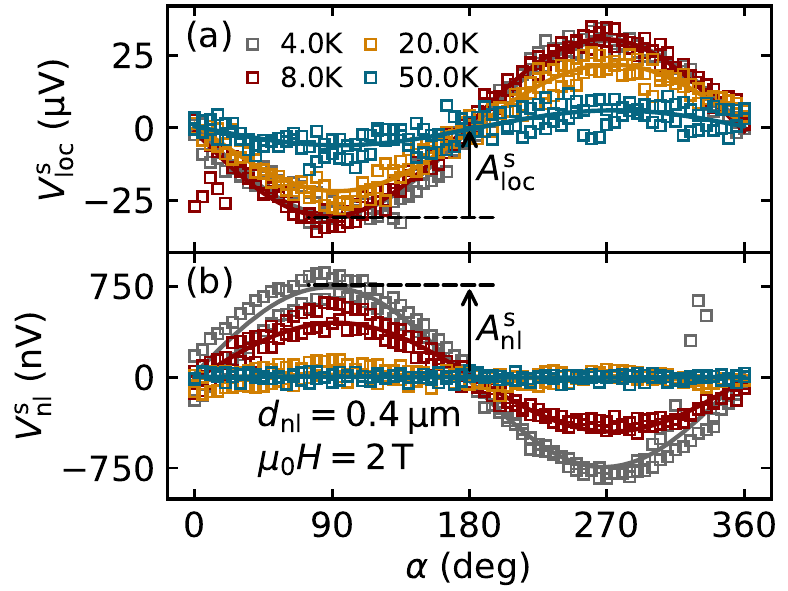}
	\caption{%
	  The angle dependence of (a) the  local SSE signal $V^s_{loc}$ and (b) the  nonlocal SSE signal $V^s_{nl}$ is shown here for several temperatures between 4 and 50 K \red{for the device with $d_{nl}$ = 0.4 $\mu$m . Here $A^s_{loc(nl)}$ indicated by arrow represent amplitude of SSE signal modulation.}  A clear increase of the amplitude of the SSE signal modulation \red{$A^s_{loc(nl)}$ } is evident towards lower temperatures. An average offset voltage independent of the magnetic field direction was subtracted from the data. 
}
	\label{AD-T}
\end{figure}

\begin{figure}[tb]
	\centering
			\includegraphics{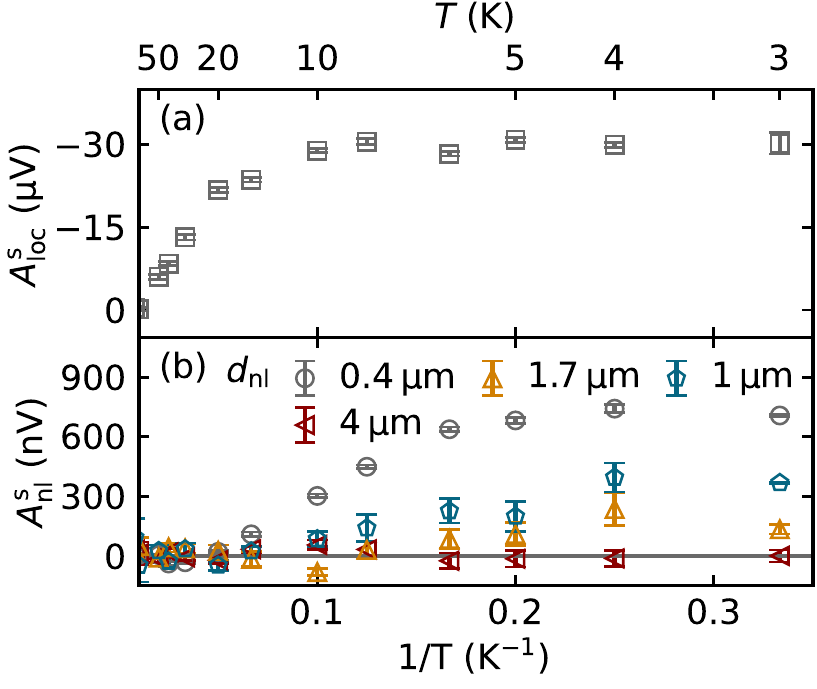}
	\caption{%
		The amplitudes of the sin$\alpha$-type modulation \red{$A^s_{loc(nl)}$} in (a) the local and (b) the nonlocal SSE voltage as a function of reciprocal temperature \red{for devices with different $d_{nl}$}. A clear increase of the signals towards lower temperatures is evident. Interestingly, the local signal $V^s_{loc}$ saturates already around 10 K, while the nonlocal signal $V^s_{nl}$ saturates only around 5 K. The measurements were done with an in-plane  magnetic field ($\alpha$-rotation) $\mu_0H$ = 2 T.
}
	\label{RNL-T}
\end{figure}

Figure 3 shows the  nonlocal SSE signal at several temperatures, measured with an applied magnetic field of $\mu_0H$ = 2 T rotated in-plane ($\alpha$ rotation). A significant change of the local and nonlocal signal magnitude as a function of temperature is evident from the data. Both the local and nonlocal signal show a stronger modulation for lower  temperature, while the modulation seems to disappear at higher temperatures. \red{The amplitude of the voltage modulation $A^s_{loc(nl)}$ represents the SSE signal without thermoelectric offsets.
Expressing our SSE signal magnitude in a nonlocal resistance (normalized to the wire length), we obtain values of  $\approx$0.23 $\frac{V}{\mu m A^2}$ in good agreement with Yuan at al.\cite{Yuaneaat1098} (see the Supplementary Material section S4 for more details)}
To better compare the temperature evolution, the amplitudes of the voltage modulations \red{$A^s_{loc(nl)}$} were extracted using a sin$(\alpha)$ fit and compiled in Fig. 4. Although both the local and the nonlocal signal saturate at low temperature, the local signal \red{$A^s_{loc}$} saturates already around 10 K, while the modulation in the nonlocal voltage \red{$A^s_{nl}$} does so  only below 5 K. Also, the modulation in the local SSE signal vanishes at around 100 K, whereas the  nonlocal SSE signal disappears already above  20 K. This suggests that although the general trends are similar, the detailed mechanisms relevant for the local and nonlocal SSE signals might differ. This can be understood considering that the thermal magnon excitation which is important for the local SSE signal, depends only on the temperature gradient $\nabla T_n$ underneath the injector, while the nonlocal signal is caused by magnon transport reaching  far beyond the thermal gradient generated by the injected electrode.

So far, the exact nature of the increase of the thermal signal at low temperatures is not fully resolved. In ref. \cite{Yuaneaat1098}, the low-temperature saturation in the nonlocal signal was attributed to a spin-superfluid ground state in the antiferromagnetic $\mathrm{Cr_2O_3}$,  resulting from spontaneous breaking of the uniaxial symmetry. However, a similar behavior was also observed in Pt-YIG-Pt lateral devices, where a spin-superfluid ground state cannot be realized\cite{Oyanagi-doi:10.1063/1.5135944}. Furthermore, in  $\mathrm{MnF_2}$ -Pt bilayers, the temperature dependence of the SSE signal shows a magnetic-field-dependent peak around T $\approx$ 7 K, which  approximately matches the peak in thermal conductivity of $\mathrm{MnF_2}$\cite{Wu-MnF2-PhysRevLett.116.097204}. In the case of $\mathrm{Cr_2O_3}$, a peak in thermal conductivity was observed for T $\approx$ 30 K\cite{Seki-Cr2O3-PhysRevLett.115.266601}. Taken together, the low-temperature SSE response in antiferromagnets is far from well understood and requires further investigation.


\begin{figure}[tb]
	\centering
			\includegraphics{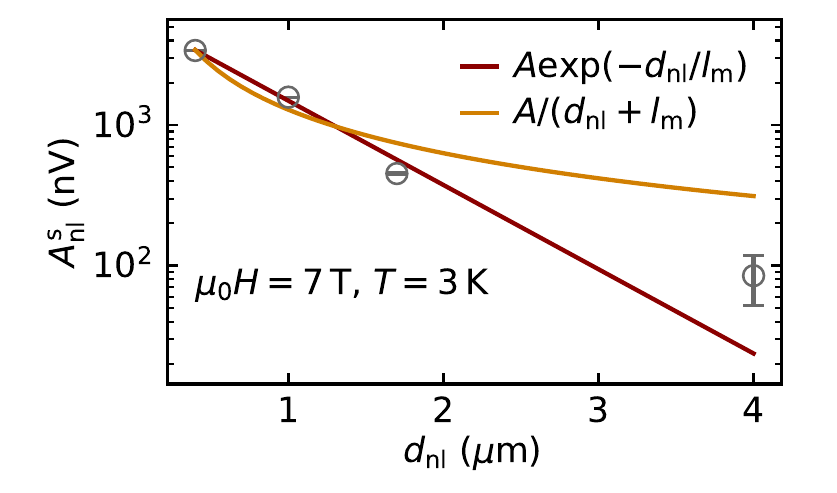} 
	\caption{
	 The nonlocal SSE signal decreases for increasing injector-detector separation $d_{nl}$. Hereby the decay is better described by an exponential, rather than an algebraic decay, as discussed in more detail in the text. The data were taken with an applied magnetic field of $\mu_0H$ = 7 T at 3 K. We estimate  a characteristic magnon decay length $l_m$ $\approx$ 500 nm for these conditions.
}
	\label{RNL-d}
\end{figure}


Finally, we measured the evolution of the nonlocal SSE signal \red{$A^s_{nl}$}as a function of the injector-detector separation $d_{nl}$,  as shown in Fig. 5. \red{Beyond $d_{nl}$ = 4 $\mu$m, we could not measure $A_{nl}^s$ accurately anymore, as the nonlocal voltage reached the noise level of the nanovoltmeter, and the signal-to-noise ratio become too small. Therefore, we focus only on the "small-distance" regime here.} The length scale for the magnon spin current can be estimated from this \red{$A^s_{nl}$} vs $d_{nl}$  data considering a one-dimensional spin diffusion model which describes the decay as\cite{Cornelissen2015} 
\red{
\begin{equation}
    A_{nl}^S  = C\exp \left( { - \frac{{d_{nl} }}{{l_m }}} \right),
\end{equation}
}
where $l_m$ is the magnon spin diffusion length and C is a constant independent of $d_{nl}$. We find this simple exponential decay of the signal (red line) fits best to the measured data, and we estimate  $l_m$ = 500 nm  at 3 K. This is quite small compared to the magnon spin diffusion length in the ferrimagnetic insulator YIG,  where $l_m$ = 40 $\mu$m at T = 3.5 K was reported using a similar nonlocal method.\cite{Shan-PhysRevB.96.184427}. This is also quite small compared to the magnon spin diffusion length in other antiferromagnets like $\mathrm{\alpha-Fe_2O_3}$ with $l_m$ = 9 $\mu$m at 200 K \cite{Lebrun2018}.
In previous spin transport experiments in $\mathrm{Cr_2O_3}$ thin films, Yuan et. al.\cite{Yuaneaat1098} reported a very large $l_m$ = 16.3 $\mu$m and assumed  a spin-superfluid ground state to be realized in $\mathrm{Cr_2O_3}$. However, short ($<~10~nm$) spin decay lengths through AFI were observed in vertical (longitudinal geometry) spin transport devices\cite{Wang-PhysRevB.91.220410}.

In magnon spin transport experiments in YIG, three distinct transport regimes have been identified. For very short distances ($d~<<~l_m$), the magnon signals typically drop faster than the exponential decay (Eq. (1))\cite{Shan-PhysRevB.94.174437,Shan-PhysRevB.96.184427}. Thereafter, for intermediate distances, an exponential decay is observed\cite{Cornelissen2015}. This regime is called the "exponential regime" or "relaxation regime". Beyond this intermediate regime, \red{$A^s_{nl}$} shows a geometrical decay as \red{$A_{nl}^S \sim\frac{1}{{d_{nl}^2 }}$}\cite{Shan-PhysRevB.96.184427}. In this geometrical-decay regime, the signal is controlled by the temperature gradient $\nabla T$ present close to the detector, which also generates a magnon spin current that contributes to the measured nonlocal SSE signal. The good match of \red{$A_{nl}^S$} to Eq.[1] shown in Fig. 5 indicates that the spin signal in our case is dominated by magnon relaxation rather than diffusive transport. 

In case of a spin-superfluid ground state, the nonlocal signal was predicted to follow a decay of the form\cite{Yuaneaat1098,Takei-PhysRevB.90.094408} 
\red{
\begin{equation}
     A_{nl}^S\sim\frac{1}{{d_{nl}  + l_m }},
\end{equation}
}
As evident from Fig. 5, in our case, the exponential decay given by  Eq. (1)  fits much better to the data than an algebraic decay given by  Eq. (2). 


The relatively short magnon relaxation length along with the strong temperature dependence observed in our experiment suggest that temperature-dependent inelastic magnon scattering processes play an important role in long-range magnon transport through antiferromagnets\cite{Troncoso-PhysRevB.101.054404,Bender-PhysRevLett.119.056804}. Note that from our previous structural and magnetic characterization of $\alpha$-$\mathrm{Cr_2O_3}$ thin films \red{(See Supplementary Material section S1)}, we have found that our films are of high crystalline quality but granular with a typical grain size of $\sim$50 nm. In addition, they contain magnetic domains of typical dimension $\sim$230 nm\cite{Kosub-PhysRevLett.115.097201,Kosub2017,Patrick-Appe-doi:10.1021/acs.nanolett.8b04681}. 
The similar  scale of magnon spin relaxation length $l_m$ and domain size suggests that inelastic magnon scattering processes from uncompensated magnetic moments at the grain boundaries or antiferomagnetic domain walls significantly impact or even  dominate magnon transport in our $\alpha$-$\mathrm{Cr_2O_3}$ thin films\cite{Tveten-PhysRevLett.112.147204,Ross-Klaui}. 
Note that also in YIG, the magnon spin diffusion length depends strongly on the crystalline quality and texture. For example, at room temperature $l_m$ = 38 nm for sputtered YIG\cite{Sputter-YIG-PhysRevB.93.060403}, $l_m$ = 140 nm for pulsed-laser-deposited YIG \cite{PLD-YIG-PhysRevLett.115.096602}and $l_m$ = 9.4 $\mu$m for liquid-phase-epitaxy YIG were observed\cite{Cornelissen2015}.


    In summary, we performed a detailed study of  thermally excited  magnon transport in antiferromagnetic $\alpha$-$\mathrm{Cr_2O_3}$ thin films using a nonlocal device geometry. No direct electronic excitation of magnons could be observed, while a clear voltage signal arising from thermally  generated magnons was picked up in the symmetric voltage. We found that the thermally generated  local and nonlocal voltages, measured while rotating a magnetic field of constant magnitude in three orthogonal planes, follow $\sin(\alpha)$ or $\sin(\beta)$ dependencies, as expected for  the SSE. Temperature-dependent measurements showed that the SSE signals increase with decreasing temperature and saturate at very low temperatures. Finally, we estimated the length scale over which the thermally generated magnons diffuse by measuring the nonlocal SSE voltage signal as a function of the spatial separation of the injector and detector Pt strips. We conclude that magnon transport in our $\alpha$-$\mathrm{Cr_2O_3}$ thin films is governed mainly by relaxation processes with a characteristic magnon spin diffusion length of $l_m$ = 500 nm. The comparatively short magnon spin diffusion length and the strong temperature dependence of both the local and nonlocal SSE signals suggest that inelastic magnon scattering processes at the grain boundary or antiferomagnetic domain walls dominate the magnon transport at short distances in our samples.  Our findings can inspire antiferromagnetic magnonic devices, such as non-volatile memory storage, logic gates, analog data processing, and quantum computing. 
\begin{center}
\textbf{SUPPLEMENTARY MATERIAL}\\
\end{center}   

See supplementary material for further details of S1. TEM characterization, S2. Device fabrication, S3. Current dependence of the nonlocal signal, S4. Thermoelectric offsets, S5. Angle dependence of the antisymmetric voltage, S6. Field dependence of the SSE signal, and S7. Temperature dependence of Pt. 

\begin{acknowledgments}
We acknowledge funding from the Würzburg-Dresden Cluster of Excellence ct.qmat (EXC 2147, Project ID 390858490). We also acknowledge financial support by the Deutsche Forschungsgemeinschaft via SFB 1143/C08, SPP 1538 [Project Nos. GO 944/4, TH1399/5], MA5144/9-1, MA5144/24-1, MA5144/22-1, and Helmholtz Association of German Research Centres in the frame of the Helmholtz Innovation Lab “FlexiSens” (HIL-A04)). Furthermore, the use of the HZDR Ion Beam Center TEM facilities is gratefully acknowledged. We also thank R. Aniol for TEM specimen preparation and T. Schönherr for technical support during ebl fabrication of devices. 
    	
\end{acknowledgments}

\begin{center}
\textbf{DATA AVAILABILITY}\\
\end{center}
The data that support the findings of this study are available from the corresponding author upon reasonable request


\end{document}